\documentclass[prb,aps,twocolumn,showpacs,bibnotes,superscriptaddress,epsf]{revtex4}

\usepackage{graphicx}
\usepackage{amsmath}
\usepackage{amssymb}
\usepackage{color}
\usepackage{dcolumn}
\usepackage{epsfig}
\usepackage{bm}
\usepackage{subfigure}
\usepackage[urlcolor=blue]{hyperref}
\hypersetup{backref, colorlinks=true, linkcolor=blue, citecolor=blue}

\begin{document}

\title{Superconductivity in potassium-doped 2,2$'$-bipyridine}

\author{Kai Zhang}
%\affiliation{Key Laboratory of Materials Physics, Institute of Solid State Physics, Chinese Academy of Sciences, Hefei 230031, China}
%\affiliation{University of Science and Technology of China, Hefei 230026, China}
\affiliation{Center for High Pressure Science and Technology Advanced Research, Shanghai 201203, China}

\author{Ren-Shu Wang}
\affiliation{Center for High Pressure Science and Technology Advanced Research, Shanghai 201203, China}
\affiliation{School of Materials Science and Engineering, Hubei University, Wuhan 430062, China}

\author{An-Jun Qin}
\affiliation{South China University of Technology, Guangzhou 501540, China}

\author{Xiao-Jia Chen}
\email{xjchen@hpstar.ac.cn}
\affiliation{Center for High Pressure Science and Technology Advanced Research, Shanghai 201203, China}

\date{\today}

\begin{abstract}
Organic compounds are always promising candidates of superconductors with high transition temperatures. We examine this proposal by choosing 2,2$'$-bipyridine solely composed by C, H, and N atoms. The presence of Meissner effect with a transition temperature of 7.2 K in this material upon potassium doping is demonstrated by the $dc$ magnetic susceptibility measurements. The real part of the $ac$ susceptibility exhibits the same transition temperature as that in $dc$ magnetization, and a sharp peak appeared in the imaginary part indicates the formation of the weakly linked superconducting vortex current. The occurence of superconductivity is further supported by the resistance drop at the transition together with its suppression by the applied magnetic fields. The superconducting phase is identified to be K$_3$-2,2$'$-bipyridine from the analysis of Raman scattering spectra. This work not only opens an encouraging window for finding superconductivity after optoelectronics in 2,2$'$-bipyridine-based materials but also offers an example to realize superconductivity from conducting polymers and their derivatives.
\end{abstract}

\maketitle

\section{Introduction}

Organic materials, the important and ubiquitous substance in human beings' lives, are always composed of hydrocarbons containing only carbon and hydrogen, as well as other elements, especially nitrogen. Doping with metals in organic molecular compounds has been extensively studied for a long time. It has raised a storm of superconductors discovery ever since the first simple carbon-based superconductor C$_8$A (A=K, Rb, or Cs), which exhibits a superconducting transition temperature ($T_c$) of 0.02-0.55 K, found in 1965.\cite{Hanna} Later, in 1980, di-(tetramethyltetraselenafulvalene)-hexafluorophosphate, (TMTSF)$_2$PF$_6$,\cite{Jerom} a real organic superconductor composed by C and H atoms, was discovered to exhibit a $T_c$ of 0.9 K at pressure of 1.2 GPa. Interestingly, the superconductivity in (TMTSF)$_2$PF$_6$ occurs with the suppression of the spin-density-wave instability at high pressures.\cite{Ardav,Vulet} Other materials, such as the organic charge-transfer salts based on donor molecules,\cite{Murat,Yagub,Willi,Wang-4,Kano} the edge shared aromatic hydrocarbons,\cite{Mitsu,Wang,Xue} also have been observed to superconduct. In addition, it is worth mentioning that $p$-terphenyl was found to be superconductors with $T_c$ as high as 123 K,\cite{Wang-3} far above liquid nitrogen boiling temperature, which could comparable to the record high $T_c$ of about 130 K in some cuprate superconductors.\cite{Putil} This discovery gives rise to great interests in finding high-$T_c$ superconductors in organic materials. Meanwhile, superconducting phases with $T_c'$s of 7.2 K and 43 K were also observed in the synthetic.\cite{Wang-1,Wang-2} $p$-Terphenyl belongs to poly(\emph{para}-phenylene) materials, which have the phenyl rings connect at the $para$ positions. Previous works\cite{Havin,Shack} reported an increasing conductivity with increasing the phenyl ring number. Hence, exploring superconductivity in poly(para-phenylene), especially in the shorter polymers, is of great importance and necessity. Once superconductivity can be realized in the shortest polymer, the other polymers with longer chain length will capture tremendous potentials to be superconductors.

Biphenyl is the shortest polymer with only two phenyl rings connected by single C-C bond. It shows a relative simple monoclinic structure, and undergoes an incommensurate triclinic transition at about 40 K.\cite{Atake} A solid isologue with biphenyl, 2,2$'$-bipyridine, also always attracts intensive interest within the community. Comparing to biphenyl, 2,2$'$-bipyridine has a replacement of the two \emph{ortho}-CH groups with nitrogen atoms.\cite{Almen} Such a typical feature makes it a good bidentate chelating ligand. 2,2$'$-Bipyridine is a basis material in photoelectric applications.\cite{Kaes} It can form many complexes with most transition metals, such as ruthenium complexes and platinum complexes. And these complexes all exhibit intense luminescence.\cite{Kaes,Tokel,Sulli,Conni} Otherwise, the ligand, [Fe(bipyridine)$_3$]$^{2+}$, is often used for the colorimetric analysis of iron ions. As the versatile applications, thus, once this cheap and pervasive material, which composes only by C, H, and N atoms, becomes a superconductor, it will be a good example to demonstrate the obtaining of organic superconductors from the pool of extensive 2,2$'$-bipyridine-based materials. Moreover, it will extremely expand the applications of this material in superconductivity after optoelectronics.

In the present work, we report the finding of superconductivity in potassium doped 2,2$'$-bipyridine. Combination of the magnetic susceptibility and electrical transport measurements confirms that this material shows a superconducting transition with the temperature of 7.2 K. Meanwhile, the chemical formula of the superconducting phase is determined to K$_3$-2,2$'$-bipyridine from the analysis of the Raman scattering spectra. It is the first time to find superconductivity in pyridine family, which is a novel structural system composed by C, H, and N atoms in organic materials. This discovery also opens a new window to discover superconductivity from optoelectronic materials.

\section{Experimental details}

High-purity potassium (K) (99$\%$) and 2,2'-bipyridine ($>$99$\%$) were purchased from Sinopharm Chemical Reagent and Alfa Aesar, respectively. The potassium was cut into small pieces and mixed with 2,2$'$-bipyridine with a mole ratio of 3:1 in a glove box with the moisture and oxygen levels less than 0.1 ppm. Then the mixture was sealed in a quartz tube under high vacuum 1$\times$10$^{-4}$ Pa and heated at temperature of 120-130 $^\circ$C for 48-72 hours. The resulting K doped 2,2$'$-bipyridine powder sample shows uniform black color compared to the pure white color of pristine 2,2$'$-bipyridine.

The $dc$ and $ac$ magnetization measurements were performed by Magnetic Property Measurement System (Quantum Design). The sample was placed into a polypropylene powder holder (Quantum Design). The $dc$ magnetic susceptibility measurements were performed in the temperature range of 1.8-300 K with a magnetic field of 20 Oe, and the $ac$ magnetic susceptibility measurements were performed in the temperature range of 1.8-20 K under an applied field with amplitude and frequency of 8 Oe and 234 Hz, respectively. The resistance measurement was carried out by Physical Property Measurement System (Quantum Design). The sample was put into a specialized container which can generate a quasi-hydrostatic condition to make the contacts between the four leads and the sample work well. Such a high pressure ensured that the sample was isolated to the external condition. Meanwhile, the pristine 2,2$'$-bipyridine doesn't denature at this pressure, so we assume that the compressed K doped 2,2$'$-bipyridine still keeps its properties at ambition condition. The whole processes were performed in the glove box. The Raman scattering experiments were measured using a 660 nm laser line with the samples sealed in quartz tubes. The excited power was less than 2 mW before a $\times$20 objective to avoid possible damage of samples.

\section{Results and discussion}

The temperature dependence of the $dc$ magnetic susceptibility ($\chi$) of K-doped 2,2$'$-bipyridine with the field cooling (FC) and zero-field cooling (ZFC) runs under magnetic field $H$ of 20 Oe is shown in Fig. 1(a). One can clearly see a sudden drop below 7.2 K in ZFC and FC runs. The drastic drop of $\chi$, which results from the well-defined Meissner effect, indicates that a superconducting transition takes place in this sample. The temperature where the sharp drop occurred is defined as the superconducting transition temperature $T_c$. The calculated shielding fraction is about 0.7$\%$. The small shielding fraction could arise from the smaller size of crystallites than the London penetration depth.\cite{Mitsu}

\begin{figure*}[tbp]
\includegraphics[width=0.98\textwidth]{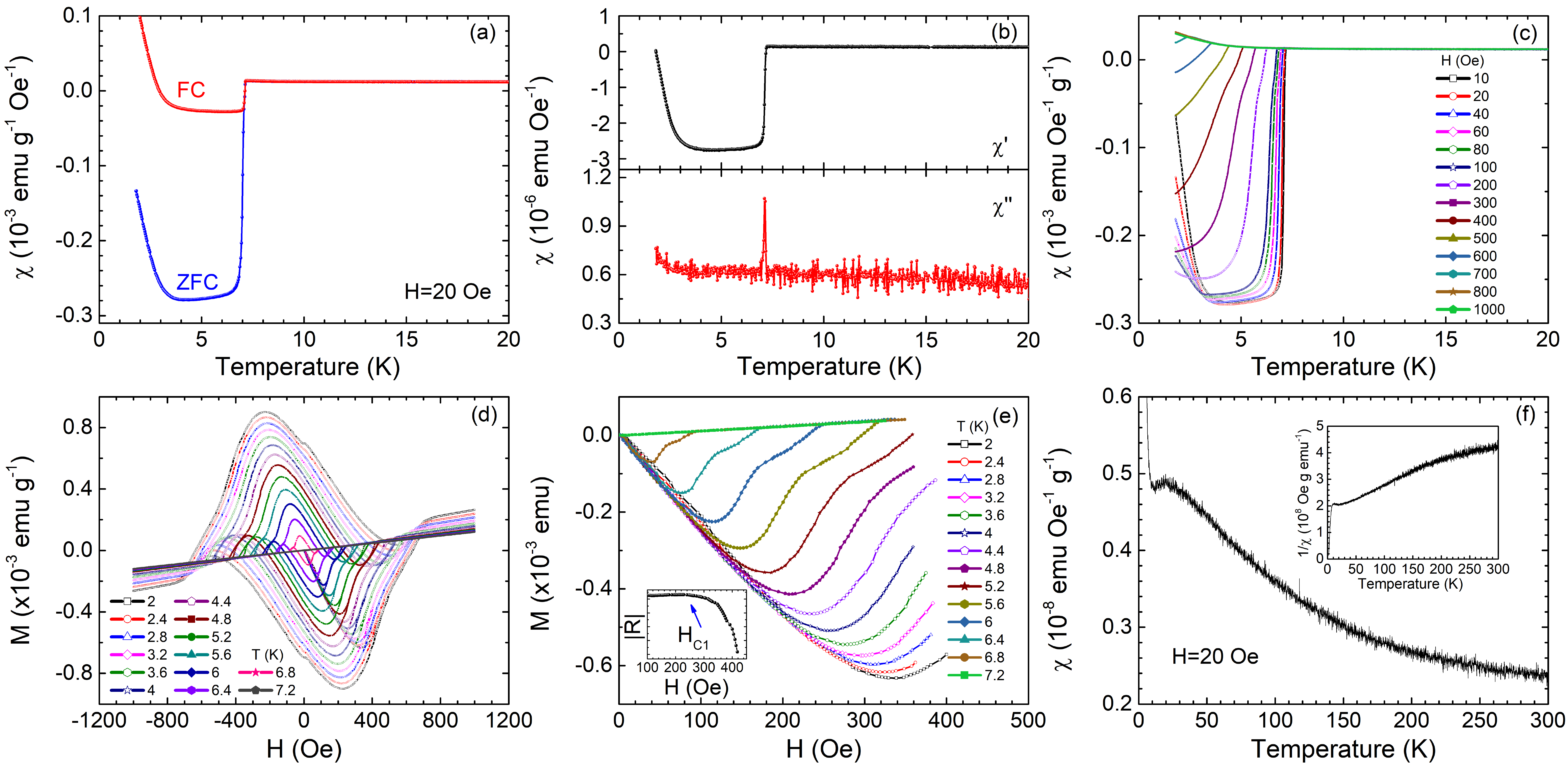}
\caption{Temperature dependence of the magnetic susceptibility $\chi$ for K-doped 2,2$'$-bipyridine. (a) Magnetic susceptibility of K$_x$bipyridine measured in the ZFC and FC runs under 20 Oe . (b) Real part and imaginary part of the $ac$ susceptibility as a function of temperature. The amplitude and frequency of applied field are 8 Oe and 234 Hz, respectively. (c) Temperature dependence of the $dc$ magnetic susceptibility of K$_x$bipyridine under various magnetic fields up to 1000 Oe in the ZFC run. (d) The isothermal hysteresis loops with scanning magnetic fields up to 1000 Oe measured at various temperatures in the superconducting state. (e) Initial part of the hysteresis loops measured at various temperatures. Inset is an example used to determine the $H_{c1}$ value. $|$R$|$ is the regression factor at 2 K. (f) Magnetic susceptibility of K$_x$bipyridine without superconducting signs in the ZFC run under 20 Oe. Inset is the reciprocal of this magnetic susceptibility.}
\end{figure*}

Figure 1(b) respectively presents the temperature dependence of the real (top) and imaginary (bottom) parts of the $ac$ magnetic susceptibility of K-doped 2,2$'$-bipyridine at 8 Oe and a frequency of 234 Hz. The real part exhibits an almost same characteristic to the $dc$ magnetic susceptibility. Meanwhile, a sharp peak appears in the imaginary part. The $ac$ measurement is a more powerful signature of the occurrence of superconductivity in this material. The complex susceptibility $\chi$=$\chi'$+i$\chi''$ is a little different from the $dc$ susceptibility. The real part $\chi'$ is a presentation of Meissner currents screening of this superconductor, and it is approximately equivalent to the ZFC susceptibility in the $dc$ measurement. The imaginary part $\chi''$ corresponds to the energy absorption in this material.\cite{Oda,Lee,Garci,Abdel-1} It arises from the formation of the weakly linked superconducting vortex current. Thus, $ac$ susceptibility can be considered as a symbol of the appearance of the superconductivity in this material. Furthermore, this parameter is also a testification of the zero resistance.\cite{Oda,Garci} Hence, our data further demonstrates the occurrence of a superconducting phase at 7.2 K in the K-doped 2,2$'$-bipyridine.

Magnetic susceptibility of K-doped 2,2$'$-bipyridine as a function of temperature in the ZFC run under different magnetic fields is also obtained. The results are depicted in Fig. 1(c). The diamagnetism of K$_x$-2,2$'$-bipyridine, which can be clearly seen in the figure, gradually decreases with increasing the magnetic field. Meanwhile, the temperature span of superconducting transition becomes significant broad as field is increased. The estimated upper critical field $H_{c2}$ is 1000 Oe, where the almost completely lose of the diamagnetic signal was observed with cooling the temperature down to 1.8 K. However, this is just a rough estimation of $H_{c2}$ because of the complicated magnetic features in the material. The reliable upper critical field obtained from resistance will be discussed in the next section.

The field dependent isothermal magnetization $M$ up to 1000 Oe at various temperatures is illustrated in Fig. 1(d). The symmetric hysteresis loops demonstrate that our material exhibits a strong bulk pinning, which is a certification of a type \uppercase\expandafter{\romannumeral2} superconductor. In addition, the slight slope of the background indicates that the sample contains magnetic impurities. These impurities should result from the part of the compound which is not superconducting. On the other hand, one can also calculate the lower critical field from the initial part of these loops.

Figure 1(e) presents the virgin $M(H)$ curves at low fields measured for the temperatures from 2 K to 7.2 K. The superconductivity is completely suppressed at 7.2 K. In order to determine the lower critical field H$_{c1}$, we calculate the regression coefficient $R$ of a linear fit to the data points. Then the $H_{c1}$ is obtained where the point in $R(H)$ starts to deviate from linear dependence. This method is enough credible which has been examined in the previous studies.\cite{Abdel,Abdel-1} The procedure used to determine $H_{c1}$ for T=2 K is shown in the inset of Fig. 1(e). Results for deduced $H_{c1}$ are shown in Fig. 2(b), depicted by yellow points. Then we use the empirical law $H_{c1}(T)/H_{c1}(0)=1-(T/T_c)^2$ to fit the lower critical field at zero-temperature $H_{c1}(0)$, the calculated $H_{c1}(0)$ is 211$\pm$30 Oe. The fitted results illustrate in Fig. 2(b) with a red line, and the superconducting area is covered with violet shade.

The synthetic non-superconducting samples always exhibit complex magnetic signals. The typical ZFC signal measured with the field of 20 Oe is shown in Fig. 1(f).  A weak peak appears at around 20 K. Reciprocal of the magnetic susceptibility was calculated to estimate the type of the magnetic signal. However, the credible type of this signal cannot be determined directly from the magnetic susceptibility measurement. Details of the type needs a further study.

\begin{figure}[tbp]
\includegraphics[width=0.48\textwidth]{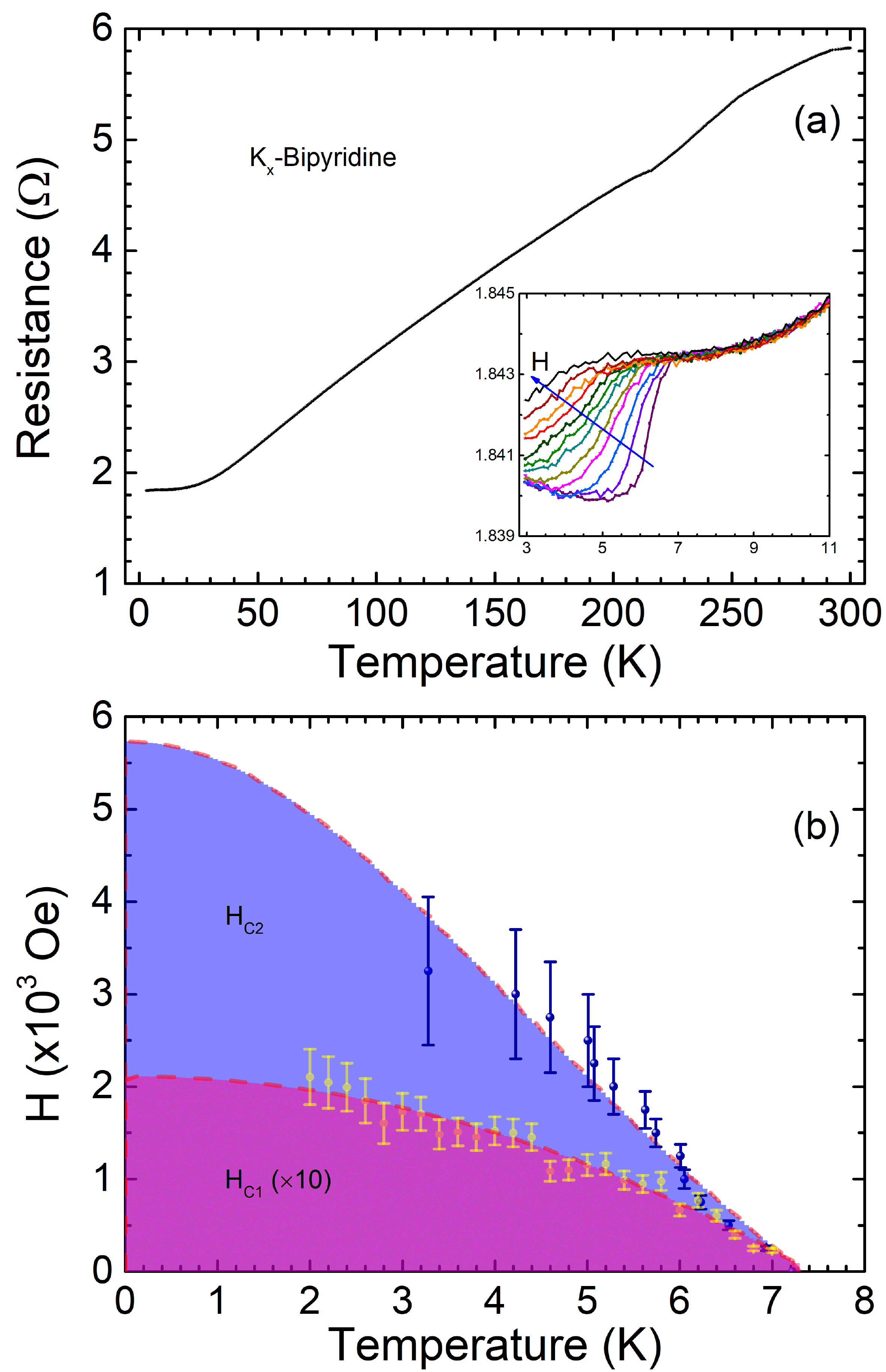}
\caption{(a) Temperature dependence of the resistance of K-doped 2,2$'$-bipyridine from 3 K to 300 K. Inset is the magneto resistance measurements under different magnetic fields. (b) The temperature dependence of the lower critical field H$_{c1}$(T) and upper critical field H$_{c2}$(T).}
\end{figure}

Zero-field resistance of K$_x$-2,2$'$-bipyridine between 3 K to 300 K is shown in Fig. 2(a). As the very low shielding fraction of this material, the resistance measurements are very hard to be duplicated. The feature of the resistance at normal state is obvious a metallic behavior. However, due to the high content of potassium, the metallic feature also possibly comes from the redundant potassium. But the resistance in the whole temperature range is much higher than pure potassium which owns a fantastic conductivity. Thus, we assume that the resistance exhibited here has the major contribution from K$_x$-2,2$'$-bipyridine. From the figure we can see that the resistance exhibits a drastic drop at around 7.2 K. This phenomenon indicates a clear superconducting transition. The temperature of 7.2 K is excellently consistent to the results of both the $dc$ and $ac$ magnetic susceptibility measurements. The inset of Fig. 2(a) shows the resistance as a function of temperature in the $dc$ magnetic fields up to 3250 Oe. Due to the damage of the sample, we cannot get the magneto resistance at higher fields. From the inset one can clearly see that the superconducting transition exhibits a pronounced broadening at high fields. And the sample still remains the superconductivity at 3250 Oe. The upper critical field $H_{c2}$ is defined by using the onset $T_c$ criteria, which is determined by the first dropped point deviated from the linear resistance curves. The results are shown in Fig. 2(b) by a blue line. We then fit the $H_{c2}(T)$ by Werthamer-Helfand-Hohenberg phenomenological formula.\cite{Werth} The fitted results are plotted in Fig. 2(b), and the superconducting area is covered with blue shade. The deduced zero-temperature $H_{c2}(0)$ is 5731$\pm$600 Oe. Combining with $H_{c1}(0)$, we evaluate the London penetration depth $\lambda$$_L$ and Ginzburg-Landau coherence length $\xi$$_{GL}$ by using the expressions\cite{Tinkh} $H_{c2}$(0)= $\Phi_{0}/2\pi\xi_{GL}^{2}$ and $H_{c1}$(0)$=(\Phi_{0}/4\pi\lambda_{L}^{2})\ln(\lambda_{L}/\xi_{GL})$. Here, $\Phi_{0}$=$h$/$e^*$=2.07$\times$10$^{-7}$ Oe cm$^2$ is the magnetic-flux quantum. The resulting $\lambda$$_L$=26$\pm$1 nm and $\xi$$_{GL}$=24$\pm$1 nm. Thus, the Ginzburg-Landau parameter $\kappa$=$\lambda$$_L$/$\xi_{GL}$=1.09$>$1/$\sqrt{2}$ is obtained for this superconductor. This result indicates that the superconductor K$_x$-2,2$'$-bipyridine is a type \uppercase\expandafter{\romannumeral2} superconductor with a transition temperature of 7.2 K.

Raman spectroscopy is a powerful tool to detect the molecular dynamics process in the organic materials. A lot of previous works\cite{Pichl,Fujik,Eklun,Kambe,Rao,Huang} focused on the structure transition of organic materials by Raman scattering measurements. Here, we present the Raman spectra of the pristine 2,2$'$-bipyridine, K-doped 2,2$'$-bipyridine, and biphenyl. The introduction of biphenyl, which shares the same crystal structure with 2,2$'$-bipyridine and $para$-terphenyl, is to compare the vibrational modes with 2,2$'$-bipyridine. The Raman spectra of 2,2$'$-bipyridine are significantly different from those of biphenyl even though they own the same monoclinic system. The distinctions should arise from the replacement of the N atoms.

The greatest distinction of these two spectra seen from the classifications of vibration modes of 2,2$'$-bipyridine\cite{Caste,Muniz,Mouss} and biphenyl\cite{Honda} present in Figs. 3(a) and 3(c) is the peaks with strong intensities range from 1200 cm$^{-1}$ to 1600 cm$^{-1}$, especially the additional peaks located around 1300 cm$^{-1}$, 1450 cm$^{-1}$, and 1485 cm$^{-1}$. Modes with frequencies located at those ranges in alkali doped poly-$para$-phenylene are often regarded as the polarons or bipolarons.\cite{Furuk} And these two elementary excitations play important roles in the conductivity of these materials. They also receive great attentions because of the assumption in the new high $T_c$ superconductors by several investigators.\cite{Alexa,Alexa-1,Jongh} Comparing the pristine and K-doped 2,2$'$-bipyridine, several features can be drawn as follows:

\begin{itemize}
  \item The lattice vibrational modes absolutely disappear and instead they merge to a broad hill.
  \item The ring breathing mode at around 1000 cm$^{-1}$ and the in-plane ring deforming modes at around 1600 cm$^{-1}$ exhibit red-shift after doping.
  \item The C-H bending and ring deforming modes at about 1200 cm$^{-1}$ to 1500 cm$^{-1}$ have the energy increases after doping.
  \item The C-H stretching bands at high frequency almost disappear after doping.
\end{itemize}

\begin{figure}[tbp]
\includegraphics[width=0.48\textwidth]{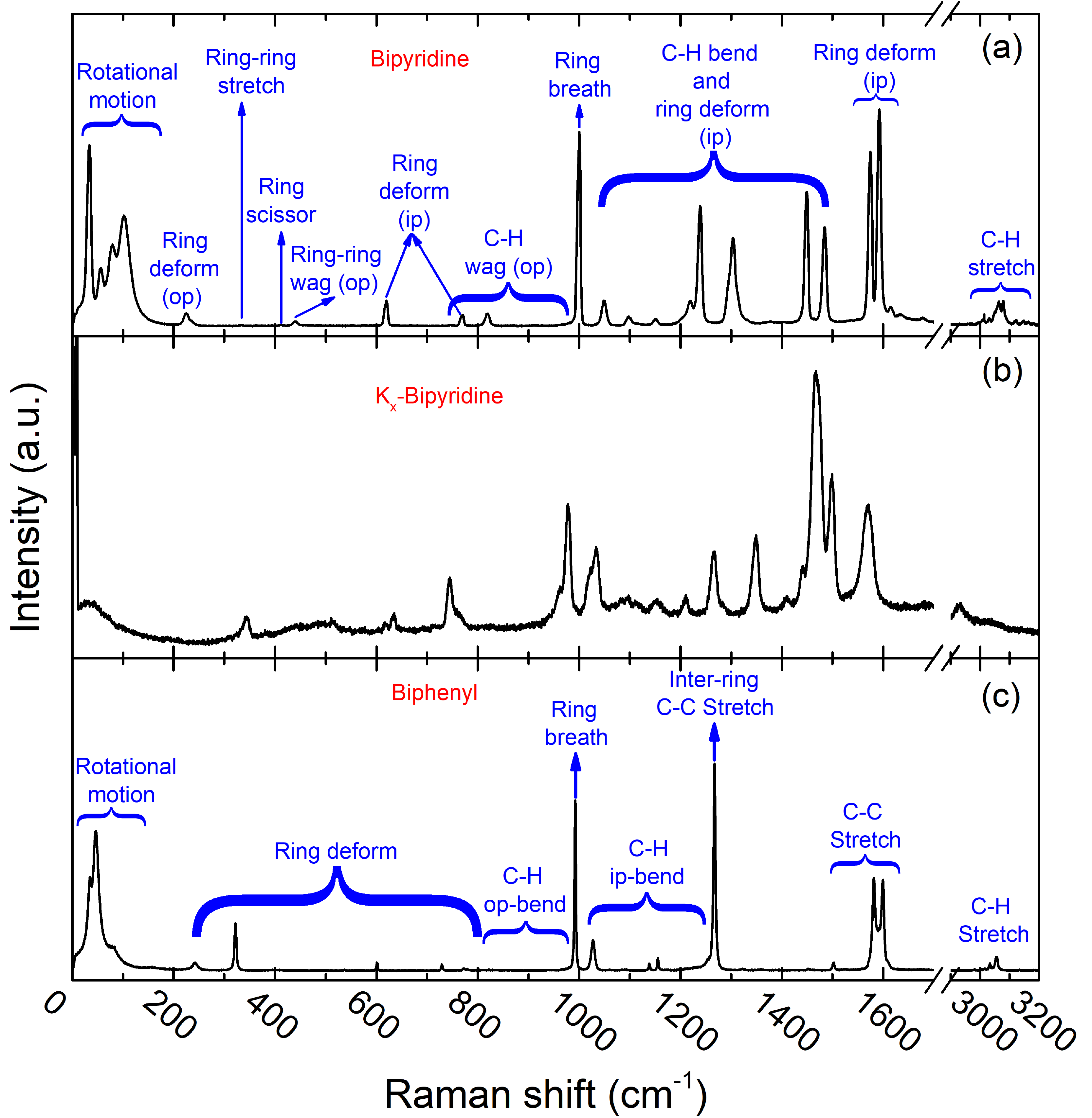}
\caption{Raman scattering spectra of pristine (a), K-doped 2,2$'$-bipyridine (b), and biphenyl (c) at room temperature. The sticks in the upper horizontal axis give the peak positions of the vibrational modes in pristine materials.}
\end{figure}

The doped potassium in 2,2$'$-bipyridine will have a significant impact on the superconducting properties. It can elongate the intralayer C-C bond length because of the intercalation between the dopants and 2,2$'$-bipyridine. On the other hand, the potassium, as an electron donator, will transfer a certain amount of electrons to 2,2$'$-bipyridine, and the number of the electrons can be detected from the Raman scattering spectra.\cite{Pichl,Fujik,Kambe,Eklun,Rao,Huang} In 2,2$'$-bipyridine, the modes at around 1600 cm$^{-1}$ are closely related to the C-C stretching. Thus, it can be chosen to calculate the amount of charge transfer. From Figs. 3(a) and 3(b) one can see that the peak with lower frequency almost keeps its energy after doping, whereas the higher frequency exhibits a 21 cm$^{-1}$ downshift from 1592 cm$^{-1}$ to 1571 cm$^{-1}$. From previous works,\cite{Huang} each electron contributes a 7 cm$^{-1}$ redshift to the Raman mode. Thus, there should be three electrons transferred to 2,2$'$-bipyridine, $i.e.$, the superconducting phase in the synthetic material is K$_3$-2,2$'$-bipyridine.

Due to the very small content of the superconducting phase and the complex components in the K-doped 2,2$'$-bipyridine, the XRD data is very hard to be analyzed.

\section{Conclusion}

Superconductivity with the transition temperature of 7.2 K has been found in potassium doped 2,2$'$-bipyridine. The occurrence of the superconducting phase has been confirmed by the combination of magnetic and electrical transport measurements. Drastic drops exhibit in the $dc$ magnetic susceptibility under the field cooling and zero-field cooling runs, as well as the real part of the $ac$ magnetic susceptibility. Meanwhile, a sharp peak has been found in the imaginary part of the $ac$ magnetic susceptibility, which indicates an existence of bulk superconductivity in the investigated compound. Temperature dependent magnetic susceptibility in the ZFC run under different magnetic fields further confirms the Meissner diamagnetism of K$_x$-2,2$'$-bipyridine. In the resistance measurement, a drastic drop, which results from the zero-resistance effect of superconductivity, was also found. The resistance is gradually suppressed with increasing the magnetic field. A phase diagram of the upper critical field and lower critical field calculated respectively from the initial part of the superconducting hysteresis loops and the magneto-resistance was built. The Ginzburg-Landau parameter $\kappa$=1.09 was obtained, this value indicates that the synthetic superconductor K$_3$-2,2$'$-bipyridine is a type \uppercase\expandafter{\romannumeral2} superconductor. The Raman spectra of the potassium doped 2,2$'$-bipyridine indicate that there should be three electrons transferred to 2,2$'$-bipyridine. Thus, we assume that the superconducting phase is K$_3$-2,2$'$-bipyridine.

The discovery of the superconductivity in potassium doped 2,2$'$-bipyridine is an interesting and significant achievement in the organic materials. It's the first time to find superconductivity in the pyridine family. It opens a new door for finding superconductivity in organic materials composed by C, H, and N atoms. Moreover, 2,2$'$-bipyridine is a basis material in photoelectric applications. This discovery extremely expands the applications in the superconducting field. In addition, 2,2$'$-bipyridine is the shortest polymer in pyridine family, the superconductivity of K$_3$-2,2$'$-bipyridine offers a potential avenue to find superconductors in the long polymers and their derivatives.

We thank Guo-Hua Zhong, Hai-Qing Lin, Yun Gao, and Zhong-Bing Huang for valuable discussions.

\end{document}